\documentclass[a4paper,10pt,fleqn]{article}
\pdfoutput=1

\usepackage{jheppub}
\usepackage[T1]{fontenc}
\usepackage{graphicx}
\usepackage{dcolumn}
\usepackage{bm}
\usepackage{simplewick}

\usepackage{tikz}
\usepackage{pgfplots}
\usepackage{microtype}

\newcommand{\ket}[1]{\ensuremath{| {#1} \rangle }}

\renewcommand{\vec}[1]{\bm{#1}}
\renewcommand{\triangle}{\Delta}
\renewcommand{\imath}{i}
\newcommand{\tr}{\text{tr}}

\title{\scalebox{.88}{Gauge invariant determination of charged hadron masses}}

\author[a]{M. Hansen,}
\author[b]{B. Lucini,}
\author[c,d]{A. Patella,}
\author[e]{N. Tantalo}
\author{\textnormal{(RC$^\star$ collaboration)}}

\affiliation[a]{CP3--Origins, University of Southern Denmark,\\Campusvej 55, DK--5230 Odense M, Denmark.}
\affiliation[b]{College of Science, Swansea University,\\Singleton Park, Swansea, SA2 8PP, UK.}
\affiliation[c]{CERN, Department of Theoretical Physics,\\CH--1211 Geneva 23, Switzerland.}
\affiliation[d]{School of Computing, Electronics and Mathematics, and Centre for Mathematical Science,\\
Plymouth University, Drake Circus, Plymouth, PL4 8AA UK.}
\affiliation[e]{University of Rome Tor Vergata, \textit{and} INFN Roma Tor Vergata,\\
Via della Ricerca Scientifica 1, I--00133, Rome, Italy.}

\emailAdd{hansen@cp3.sdu.dk}
\emailAdd{b.lucini@swansea.ac.uk}
\emailAdd{agostino.patella@cern.ch}
\emailAdd{nazario.tantalo@roma2.infn.it}

\preprint{CP3-Origins-2018-006-DNRF90, CERN-TH-2018-033}

\abstract{
In this paper we show, for the first time, that charged--hadron masses can be calculated on the lattice without relying on gauge fixing at any stage of the calculations. In our simulations we follow a recent proposal and formulate full QCD$+$QED on a finite volume, without spoiling locality, by imposing C--periodic boundary conditions in the spatial directions. Electrically charged states are interpolated with a class of operators, originally suggested by Dirac and built as functionals of the photon field, that are invariant under local gauge transformations. We show that the quality of the numerical signal of charged--hadron masses is the same as in the neutral sector and that charged--neutral mass splittings can be calculated with satisfactory accuracy in this setup. We also discuss how to describe states of charged hadrons with real photons in a fully gauge--invariant way by providing a first evidence that the proposed strategy can be numerically viable.
}

\begin{document} 
\maketitle
\flushbottom

\section{Introduction}
QED radiative corrections to hadronic observables are generally rather small but they become phenomenologically relevant when the target precision is at the percent level. For example, hadron masses and leptonic decay rates of light pseudoscalar mesons are among the best measured hadronic observables and they have to be calculated at the same level of precision. Presently, these quantities can be calculated with  percent accuracy by performing lattice simulations of QCD$+$QED, see e.g. refs.~\cite{Giusti:2017dwk,Boyle:2017gzv,Giusti:2017dmp,Fodor:2016bgu,Endres:2015gda,Borsanyi:2014jba} for a selection of recent papers on the subject and refs.~\cite{Tantalo:2013maa,Portelli:2015wna,Patella:2017fgk} for recent reviews. All these calculations have been performed by using non--compact gauge--fixed lattice formulations of QED in a finite box, see ref.~\cite{Patella:2017fgk}.    

In ref.~\cite{Lucini:2015hfa} it has been argued that charged--hadron masses can be calculated on the lattice from first principles, in a completely gauge--invariant setup, without spoiling basic QFT principles in finite volume, in particular locality.
This result is far from obvious. The construction is possible thanks to two crucial ingredients: a slightly unconventional compact formulation of lattice QED, and properly chosen boundary conditions in the spatial directions.

In a gauge theory physical states are invariant under local gauge transformations. Therefore, in order to avoid gauge fixing, physical states have to be probed by using interpolating operators that are invariant under local gauge transformations. Building these operators is trivial in the neutral sector of the theory. For example, in order to compute the mass of a neutral kaon one can use $\bar s \gamma_5 d$ as the interpolating operator. Since the down and strange quarks have the same electric charge the operator is electrically neutral and invariant under both local and global U(1) gauge transformations. 

Remarkably, in infinite volume, one can build interpolating operators that are invariant under local gauge transformations also in the charged sector of the theory. The existence of these operators was first pointed out by Dirac in an illuminating classic paper~\cite{Dirac:1955uv} (see section~\ref{sec:diracop}). In principle, Dirac's interpolating operators can be used to calculate observables associated with charged particles, e.g. the mass of the electron or of a charged kaon, in a fully gauge--invariant way. In practice, in order to obtain a fully gauge--invariant formulation of QCD$+$QED one has to provide a regularisation of the theory where Dirac's construction can be implemented without any theoretical ambiguity.  

Dirac's construction cannot be implemented on the periodic torus. In operatorial formalism, the generator of local gauge transformations is $\partial_k E_k - j_0$, where $E_k$ is the electric field and $j_0$ is the charge density, such that $Q=\int d^3x\, j_0$. Identifying physical states, $\ket{\Psi}$, with gauge--invariant states is equivalent to requiring that physical states must satisfy the Gauss law. In particular this implies that $Q\ket{\Psi}= \int d^3x\, \partial_k E_k\ket{\Psi}$. Therefore, with periodic boundary conditions in space the global constraint imposed by the Gauss law forbids states with non--zero charge. Equivalently, no interpolating operator exists on the periodic torus which is electrically charged and invariant under local gauge transformations. 

In ref.~\cite{Lucini:2015hfa} it has been proposed to discretise QCD$+$QED on a finite lattice by using the compact formulation and, as first suggested in refs.~\cite{Kronfeld:1990qu,Kronfeld:1992ae,Wiese:1991ku,Polley:1993bn}, with C--periodic (or C$^\star$) boundary conditions in space. A detailed theoretical analysis of the theory, called QCD$+$QED$_\text{C}$, has shown that the Gauss law implies a less restrictive global constraint in this case. Some electrically charged states can be probed by implementing Dirac's original construction in a fully consistent theoretical setup (see section~\ref{sec:cstar}), i.e. by using charged interpolating operators which are invariant under local gauge transformations. 

While the theoretical analysis of ref.~\cite{Lucini:2015hfa} opens the attractive possibility to perform first--principles non--perturbative lattice simulations of QCD$+$QED in a fully gauge--invariant setup, no evidence was provided concerning the numerical viability of the proposal\footnote{
The numerical effectiveness of the gauge--invariant construction of ref.~\cite{Lucini:2015hfa} has been investigated in the context of the abelian Higgs model in ref.~\cite{Woloshyn:2017rhe} with rather satisfactory numerical results. Here the issue is addressed, for the first time, in the more realistic and phenomenologically relevant case of full QCD$+$QED lattice simulations.}. In this paper we make a first step in the direction of filling this gap. We provide clear numerical evidence that charged--hadron masses can be effectively calculated in QCD$+$QED$_\text{C}$ from the gauge--invariant interpolating operators with the same signal--to--noise ratio as their neutral almost--degenerate counterparts. We also discuss how to describe states of charged hadrons with real photons in a fully gauge--invariant way. On the other hand, the cost of the generation of configurations will be analysed in future work.

The paper is organised as follows. In section~\ref{sec:diracop} we review Dirac's original construction of gauge--invariant interpolating operators for charged states. In section~\ref{sec:cstar} we recall the finite--volume formulation of QCD$+$QED with C$^\star$ boundary conditions of ref.~\cite{Lucini:2015hfa} and the lattice construction of gauge--invariant electrically--charged operators. In section~\ref{sec:numerics} we present our numerical results for charged and neutral meson masses both in the vector and pseudoscalar channel. In particular, in subsection~\ref{sec:numericsA} we discuss the implementation of a strategy to probe charged--hadron states with real photons. We draw our conclusions in section~\ref{sec:conclusions}. Finally, in appendix~\ref{sec:gftpf} we discuss some of the subtleties arising in the charged sector when the U(1) gauge is fixed, and in appendix~\ref{sec:correlators} we provide some technical details concerning the numerical evaluation of the correlators used in this study.

\section{Dirac's interpolating operator}
\label{sec:diracop}

Dirac~\cite{Dirac:1955uv} has shown that charged states in infinite--volume QED can be described in a fully gauge--invariant setup in terms of physical degrees of freedom. In Dirac's original construction the state of an electron can be interpolated by means of the operator
\begin{gather}
\Psi_e^\text{c}(x) = 
\exp\left\{
-\imath\int d^3y\,  \Phi(\vec x-\vec y)\, \partial_k A_k(x_0,\vec y)
\right\}\, \psi_e(x) \ ,
\end{gather}
where $k$ is a spatial index, $A_\mu(x)$ and $\psi_e(x)$ are the photon and electron fields while $\Phi(\vec x)$ is the electrostatic potential satisfying 
\begin{gather}
\partial_k\partial_k \Phi(\vec x)= \delta^3(\vec x) \ .
\label{eq:poisson}
\end{gather}
Under a gauge transformation $\lambda(x)$ the fundamental fields transform as
\begin{gather}
   A_\mu(x) \to A_\mu(x) + \partial_\mu \lambda(x) \ , \qquad
   \psi_e(x) \to \exp\{\imath \lambda(x)\}\psi_e(x) \ .
\end{gather}
If $\lambda(x)$ has compact support, the integral appearing in the definition of $\Psi_e^\text{c}(x)$ transforms as
\begin{gather}
\int d^3y\,  \Phi(\vec x-\vec y)\, \partial_k A_k(x_0,\vec y)
\ \to \ \int d^3y\,  \Phi(\vec x-\vec y)\, \partial_k A_k(x_0,\vec y) + \lambda(x)
\ .
\end{gather}
The operator $\Psi_e^\text{c}(x)$ is invariant under local gauge transformations, but transforms non--trivially under global gauge transformations. When acting on the vacuum, $\Psi_e^\text{c}(x)$ generates a physical state (i.e. invariant under local gauge transformations) with total charge different from zero. 

An important observation concerning this construction is that in Coulomb gauge $\partial_k A_k(x)=0$ the interpolating operator is identically equal to $\psi_e(x)$. On the one hand, this means that Dirac's construction can be circumvented and that the mass of the electron can be calculated in Coulomb gauge by using $\psi_e(x)$ as interpolating operator. This is presumably the reason why Dirac's paper went almost forgotten. On the other hand, Dirac's construction explains \emph{why} gauge--invariant physical quantities can be conveniently extracted by working at fixed gauge.

The gauge--invariant language is very useful in order to identify and clarify some of the subtleties arising with commonly used gauge--fixing conditions. For instance the Landau--gauge elementary field $\psi_e(x)$ is identical to the following generalisation of Dirac's original operator,
\begin{gather}
\Psi_e^\ell(x) = 
\exp\left\{
\imath\int d^4y\,  \Phi_\ell(x-y)\, \partial_\mu A_\mu(y)
\right\}\, \psi_e(x) \ ,
\qquad
\partial_\mu \partial_\mu \Phi_\ell(x) = \delta^4(x) \ .
\label{eq:landau}
\end{gather}
This implies for the two--point function
\begin{gather}
   \langle \psi_e(x) \bar{\psi}_e(0) \rangle_\text{Landau gauge}
   =
   \langle \Psi^\ell_e(x) \bar{\Psi}^\ell_e(0) \rangle_\text{gauge invariant}
   \ .
\end{gather}
Since $\Psi_e^\ell(x)$ is non--local in time, a standard interpretation as an interpolating operator is not possible. The phase in eq.~\eqref{eq:landau} should rather be viewed as a term in the action. Since the term is linear in the electromagnetic field, this is in fact the coupling to a non--real external electromagnetic current.

This mechanism is quite general. As discussed in appendix~\ref{sec:gftpf}, gauge fixing introduces (except special cases, of which Coulomb gauge is the most notable one) a violation of the Gauss law in the sector of non--zero charge, which can be interpreted as the effect of coupling the physical system to an external electromagnetic four--current.
This current, and consequently the Hamiltonian, is generally time dependent. In Euclidean spacetime, as an effect of the Wick rotation, the external charge density is real while the external current density is imaginary and the Hamiltonian turns out to be non--hermitean.
This implies that a spectral decomposition of two--point functions as a sum of exponentials of the form $\sum_n a_n \exp (- t E_n)$ is simply incorrect. For reasonable enough gauges (e.g. covariant gauges) the external four--current vanishes asymptotically far away from the interpolating fields in the two--point function, and the long--distance behaviour of the two--point function is dictated by the ground state of the physical Hamiltonian, i.e. in absence of the external four--current. However, in a setup in which observables are not expanded in powers of $\alpha_\text{em}$, it is not obvious at all how to extract excited physical states, such as the finite--volume counterparts of states of charged hadrons with real photons. A gauge--invariant construction of $n$--point functions becomes of utmost relevance precisely when excited states are of interest. Because the gauge--invariant Hamiltonian is hermitean and time independent, standard spectral theory applies, and gauge invariance ensures that only physical states (i.e. states that satisfy the Gauss law) propagate at any intermediate time.

In ref.~\cite{Lucini:2015hfa} Dirac's construction has been used to provide a theoretically consistent definition of electrically charged states in a \emph{finite volume} within the framework of local field theory, as we will review in the next section.

\section{Charged states in finite volume}
\label{sec:cstar}
The formulation of QCD$+$QED$_\text{C}$ has been discussed in ref.~\cite{Lucini:2015hfa} together with a detailed analysis of its symmetries and an analytical calculation of the leading finite volume effects on the masses of charged hadrons. Here, in order to make the paper self--contained, we briefly discuss the compact lattice formulation of the theory.

Gauge degrees of freedom are encoded in the link variables $U_\mu(x)\in \text{U(1)}$ and $V_\mu(x)\in \text{SU(3)}$. All the fields obey C$^\star$ boundary conditions along the spatial directions, namely
\begin{alignat}{3}
&
U_\mu(x+\vec{\hat k} L)=U_\mu(x+\vec{\hat k} L)^* \ ,
& \qquad &
\psi_f(x+\vec{\hat k} L) = C^{-1}\bar \psi_f^T(x) \ ,
\nonumber \\
&
V_\mu(x+\vec{\hat k} L)=V_\mu(x+\vec{\hat k} L)^* \ ,
& \qquad &
\bar \psi_f(x+\vec{\hat k} L) = - \psi_f^T(x)C \ ,
\end{alignat}
where $\psi_f$ are the quark fields, $f$ is the flavour index and $C$ is the charge--conjugation matrix\footnote{The charge--conjugation matrix $C$ acts on spinor indices and it can be any invertible matrix with unit determinant such that $C \gamma_\mu C^{-1} = - \gamma_\mu^T$ where $\gamma_\mu$ are the hermitean Euclidean Dirac matrices. In four dimensions such a matrix exists and satisfies $C^T=-C$ and $C^\dagger=C^{-1}$.}. We have simulated the theory by imposing periodic boundary conditions in time.

The spatial boundary conditions for the gauge fields are imposed in a completely straightforward way. However, since C$^\star$ boundary conditions mix $\psi$ and $\bar{\psi}$, the Dirac operator $D_f$ cannot be defined as an operator acting on the space of the fields $\psi$ only, but it has to be thought as an operator acting on the quark--antiquark doublet
\begin{gather}
   \eta_f =
   \begin{pmatrix}
      \psi_f \\
      C^{-1}\bar \psi_f^T
   \end{pmatrix}
   \label{eq:eta}
   \ ,
\end{gather}
which satisfies the following boundary condition
\begin{gather}
   \eta_f(x+\vec{\hat k} L)
   =
   \sigma_1 \eta_f(x)
   \label{eq:bceta}
   \ ,
\end{gather}
where the Pauli matrix $\sigma_1$ acts on the quark--antiquark components. An explicit expression for the Dirac operator $D_f$ will be given at the end of this section.

Once the fermions are integrated out, the lattice--discretised path--integral measure turns out to be
\begin{gather}
   [dU][dV] \ e^{-S_\text{g}-S_\gamma} \prod_{f=u,d,s} \text{Pf}\, (C \sigma_1 D_f) \ ,
\end{gather}
where $S_\text{g}$ and $S_\gamma$ are the SU(3) and U(1) gauge actions respectively, $\text{Pf}$ denotes the Pfaffian, and we choose to have three dynamical quarks for definiteness. The Pfaffian is proven to be real at finite lattice spacing, and positive in the continuum limit (see appendix D in~\cite{Lucini:2015hfa}). The probability to find a negative value is expected to be negligible in our simulations with fairly heavy quarks. Therefore, we have simulated the absolute value of the Pfaffian, and monitored that the lowest eigenvalue stays significantly away from zero.

For the SU(3) gauge action $S_\text{g}$ we use the L\"uscher--Weisz discretisation~\cite{Luscher:1984xn}, while the U(1) gauge action is defined as
\begin{gather}
S_\gamma = \frac{18}{e_0^2}\sum_{x,\mu\nu} \left\{1-U_{\mu\nu}(x) \right\} \ ,
\label{eq:sgamma}
\end{gather}
where $e_0$ is the bare electric charge of the positron and $U_{\mu\nu}(x)$ is the U(1) gauge plaquette, i.e.
\begin{gather}
U_{\mu\nu}(x)=U_\mu(x)U_\nu(x+\hat \mu)U_\mu(x+\hat\nu)^{-1}U_\nu(x)^{-1} \ .
\end{gather}

The point to be noticed in previous formulae is the unconventional normalisation of the $\text{U(1)}$ gauge action, namely the factor $18/e_0^2$ instead of $1/2e_0^2$. The canonically--normalised continuum action is obtained by setting
\begin{gather}
 U_\mu(x) = \exp\left\{-\frac{\imath}{6} \int_0^a d s \, A_\mu(x+s\hat{\mu})\right\} \ .
\end{gather}
To be consistent with this normalisation, the covariant derivatives acting on the quark fields are defined with the $6q_f$--power of the $\text{U(1)}$ gauge links, where $q_f$ is the charge of $\psi_f$ in units of $e_0$. For example the forward covariant derivative acting on the flavour $f$ is given by
\begin{gather} 
\nabla^f_\mu \psi_f(x) = \frac{U_\mu(x)^{6q_f}V_\mu(x)\psi_f(x+\hat \mu)-\psi_f(x)}{a} \ .
\end{gather}
The peculiar normalisation of $S_\gamma$ is due to the fact that quarks have fractional electric charges, $q_{u,c,t}=2/3$ and $q_{d,s,b}=-1/3$, and to the fact that with this choice Dirac's interpolating operators can be discretised using analytical functions of the link variables. In the lattice formulation one can choose
\begin{gather}
\Psi_{f}^{\text{s}}(x)= 
\frac{1}{3} \sum_{k=1}^3 \psi_f(x) \prod_{s=0}^{L-1} U_k(x+s a \hat k)^{-3q_f}\ .
\label{eq:psistring-1}
\end{gather}
The $k$--th term in the sum above is the unique $\text{U(1)}$ gauge--invariant extension of the quark field in axial gauge\footnote{Even though it is not obvious, one can prove that this gauge condition can always be imposed if $k$ is a C$^\star$ direction.} $U_k(x)=1$.
The corresponding expression in the finite--volume continuum theory is
\begin{gather}
\Psi_{f}^{\text{s}}(x)
\overset{a \to 0}{=}
\frac{1}{3} \sum_{k=1}^3 \psi_f(x)
\exp\left\{\frac{\imath q_f}{2}\int_0^Lds\, A_k(x+s\hat k)\right\}\ .
\label{eq:psistring-2}
\end{gather}
Notice that, given the normalisation of $S_\gamma$, only integer powers of the link variables appear in the expression of $\Psi_{f}^{\text{s}}(x)$. One can easily prove that the operators in eqs.~\eqref{eq:psistring-1} and~\eqref{eq:psistring-2} are invariant under local U(1) gauge transformations with contractible domains, while they transform non--trivially under the residual $\mathbb{Z}_2 \subset U(1)$ global gauge symmetry (see~\cite{Lucini:2015hfa} for more details).
Under local SU(3) gauge transformations the operators in eqs.~\eqref{eq:psistring-1} and~\eqref{eq:psistring-2} transform in the same way as the elementary field $\psi_f(x)$.
Finally the sum over the direction of the string ensures that they transform under discrete spatial rotations around the point $x$ in the same irreducible spinorial representation of the dihedric group as the elementary field $\psi_f(x)$.

A discretization of Dirac's original interpolating operator, i.e. the one corresponding to Coulomb gauge, can be obtained by considering 
\begin{gather} 
A^\text{c}_\mu(x) = \triangle^{-1} \bar{\nabla}_k F_{k\mu}(x) \ ,
\label{eq:Ac}
\end{gather} 
where $\nabla_k$ and $\nabla_k^*$ are the free forward and backward lattice derivatives, $\bar \nabla_k=(\nabla_k+\nabla_k^*)/2$, $\triangle =\nabla_k \nabla^*_k$, and $F_{\mu\nu}$ is a discretisation of the U(1) field tensor. In this work we have used the standard clover discretisation for the field tensor. Notice that $A^\text{c}_\mu$ is a gauge--invariant discretisation of the photon field in Coulomb gauge, $\bar \nabla_k A_k^\text{c}=0$. In the formal continuum limit
\begin{gather} 
A^\text{c}_\mu(x)
\overset{a \to 0}{=}
\triangle^{-1} \partial_k\left\{ \partial_k A_\mu(x) - \partial_\mu A_k(x) \right\}
= 
A_\mu(x) - \int_{L^3} d^3y\, \Phi(\vec x-\vec y)\partial_\mu \partial_k A_k(t_x,\vec y)\ ,
\end{gather} 
where $\Phi(\vec x)$ is the unique electrostatic potential on the finite volume with antiperiodic boundary conditions.
Therefore,
\begin{gather}
\Psi_{f}^{\text{c}}(x)
=
\frac{1}{3} \sum_{k=1}^3
\psi_f(x) \prod_{s=0}^{L-1} \left\{ U_k(x+s a \hat k)^{-3q_f}
e^{\imath 3q_f \, A^\text{c}_k(x+s a \hat k)} \right\}
\label{eq:psicoulomb}
\end{gather}
is a consistent discretisation of Dirac's interpolating operator.

In our numerical calculations, we used both the string operator $\Psi_f^\text{s}$ and the Coulomb operator $\Psi_f^\text{c}$. Fully gauge--invariant interpolating operators for charged hadrons can be obtained by starting from the usual expressions, e.g. $\bar s \gamma_5 u$, and by replacing the quark fields with the chosen Dirac's interpolating operator, e.g. $\bar S^{\text{c}} \gamma_5 U^{\text{c}}$.

Before closing this section we give the explicit expression of the $O(a)$--improved Wilson--Dirac operator used in our simulations
\begin{align} 
D_f
=
m_{0,f} 
&+ \frac{1}{2}\sum_{\mu=0}^3
\left\{
\gamma_\mu\left( \nabla^f_{\mu} + \nabla_{\mu}^{f*} \right)
- \nabla_{\mu}^{f*}\nabla_{\mu}^f
\right\}
+ \nonumber \\
&- \frac{1}{4}\sum_{\mu\nu}
\sigma_{\mu\nu}\left\{
c_{\text{sw},f}^\text{QCD}
\begin{pmatrix}
   G_{\mu\nu} & \\
   & -G_{\mu\nu}
\end{pmatrix}
+
q_f c_{\text{sw},f}^\text{QED}
\begin{pmatrix}
   F_{\mu\nu} & \\
   & -F_{\mu\nu}
\end{pmatrix}
\right\}\;.
\label{eq:dirac}
\end{align}
The forward derivative acts on the quark--antiquark doublet $\eta_f$ as
\begin{gather} 
a \nabla^f_\mu \eta_f(x) =
\begin{pmatrix}
   U_\mu(x)^{6q_f}V_\mu(x) & \\
   & U_\mu(x)^{-6q_f}V_\mu(x)^*
\end{pmatrix}
\eta_f(x+\hat \mu)-\eta(x) \ ,
\end{gather}
and is defined at the boundary by means of the relation~\eqref{eq:bceta}. The backward derivative $\nabla_{\mu}^{f*}$ is defined analogously. $G_{\mu\nu}$ and  $F_{\mu\nu}$ are the clover discretisations of the SU(3) and U(1) field tensors respectively, and $\sigma_{\mu\nu}=\imath[\gamma_\mu,\gamma_\nu]/2$. The field tensors are normalised in such a way that tree--level improvement is achieved by choosing $c_{\text{sw},f}^\text{QCD} = c_{\text{sw},f}^\text{QED} = 1$.

\section{Numerical explorations}
\label{sec:numerics}
In this section we discuss some exploratory simulations of QCD+QED with C$^\star$ boundary conditions. The main goal of this study is to show that the masses of charged mesons can be extracted in a completely gauge invariant way, with the same quality of the numerical signal as for neutral mesons. A preliminary calculation of excited states that would correspond to states of charged mesons with one real photon at $\alpha_\text{em}=0$ is also shortly presented. 

The simulations have been performed by using 
a modified version of the \texttt{HiRep} code~\cite{DelDebbio:2008zf} (see ref.~\cite{Hansen:2017zly} for more details concerning the implementation) and we have checked our results by performing dedicated runs with the publicly--available \texttt{openQ*D} code~\cite{rcstarhome} developed independently within the RC$^\star$ collaboration (see ref.~\cite{Campos:2017fly}). While the \texttt{HiRep} code has been preferred in this exploratory work because of its simplicity, the optimized \texttt{openQ*D} code is currently used by the RC$^\star$ collaboration to perform realistic QCD+QED simulations.

We have generated two SU(3)$\times$U(1) ensembles which differ only for the electromagnetic coupling, one with $\alpha_\text{em}=1/137$ and one with $\alpha_\text{em}=0.05=6.85/137$. The lattice is $48 \times 24^3$ with periodic boundary conditions in time and C$^\star$ boundary conditions in all spatial directions. The L\"uscher--Weisz action and the action in eq.~\eqref{eq:sgamma} have been used for the SU(3) and U(1) gauge fields respectively. Three dynamical Wilson fermions with Dirac operator given in eq.~\eqref{eq:dirac} have been simulated, one up--type quark with charge $q=2/3$ and two down--type quarks with $q=-1/3$.
The QCD bare parameters have been taken from one of the $N_f=2+1$ CLS ensembles at the symmetric point, i.e. the H200 ensemble in ref.~\cite{Bruno:2014jqa} with $\beta=3.55$, $\kappa=0.137$, $c_{\text{sw},\star}^\text{QCD}=1.824865$, and complemented with the tree--level value $c_{\text{sw},\star}^\text{QED}=1$. The values in physical units of the lattice spacing and of the pseudoscalar meson masses are given in table~\ref{tab:parameters}. 

In order to obtain a similar physics in the QCD and QCD+QED ensembles, the bare parameters would need to be retuned. In particular the bare masses of the up and down quarks should be retuned separately. However for sake of simplicity, in these exploratory simulations we chose to keep the bare parameters fixed and to measure the QED effects on the physical quantities. In particular we observe that QED corrections on the lattice spacing are fairly small even at the larger value of $\alpha_\text{em}$. The effect on the critical bare mass is general larger, as expected since this is an ultraviolet divergent quantity. Nevertheless we observe that in our ensemble with $\alpha_\text{em}=1/137$ the pseudoscalar mesons have reasonable masses, of the order of the physical kaon mass.

Our simulations use a volume that is smaller than the original CLS ensemble. This is potentially an issue since masses in QCD+QED have finite volume corrections that decay as inverse powers of $L$ rather than exponentially. An estimate of the finite-volume effects can be obtained by calculating the universal $1/L$ and $1/L^2$ corrections (see sec. 5 in~\cite{Lucini:2015hfa}), which turn out to be well below 1\% for both values of $\alpha_\text{em}$.


\begin{table}

\centering

\renewcommand{\arraystretch}{1.2}
\setlength{\tabcolsep}{4mm}

\begin{tabular}{ccccc}
   $\alpha_\text{em}$ & $t_0/a^2$ & $a$ & $M_P^0$ & $M_P^\pm$ \\
   \hline
   $0$ & $5.150(25)$ & $0.064\text{ fm}$ & $420\text{ MeV}$ & $420\text{ MeV}$ \\
   $1/137$ & $4.903(39)$ & $0.066\text{ fm}$ & $460\text{ MeV}$ & $510\text{ MeV}$ \\
   $0.05$ & $3.823(22)$ & $0.075\text{ fm}$ & $660\text{ MeV}$ & $860\text{ MeV}$ \\
\end{tabular}

\caption{
Values of $t_0$ in lattice units, and approximative values for the lattice spacing and pseudoscalar meson masses in physical units. The values at $\alpha_\text{em}=0$ are the CLS ones (H200 ensemble), and have been taken from~\cite{Bruno:2014jqa}. The three ensembles share the same value of $\beta=3.55$ and $\kappa_\text{u}=\kappa_\text{d}=\kappa_\text{s}=0.137$. For our simulations at $\alpha_\text{em} \neq 0$ we use $q_\text{u}=2/3$ and $q_\text{d}=q_\text{s}=-1/3$. The lattice spacing has been estimated by rescaling the CLS value with our measured $a/\sqrt{t_0}$, and the error is estimated to be of order $10^{-3}\text{ fm}$. The error on our pseudoscalar masses is estimated to be of order $15\text{ MeV}$.
\label{tab:parameters}
}

\end{table}
   

\subsection{Charged and neutral mesons}
\label{sec:numericsA}

With C$^\star$ boundary conditions the eigenstates of the momentum are also eigenstates of charge conjugation. In particular zero--momentum states are also even under charge conjugation. The boundary conditions break the U(1) global gauge symmetry down to its $\mathbb{Z}_2$ subgroup. As a consequence, if $Q$ is the electric charge operator, then $Q$ is not conserved, but $(-1)^Q$ is. When we talk about neutral states we really talk about states with $(-1)^Q=+1$, and when we talk about charged states we really talk about states with $(-1)^Q=-1$.

We consider the following C--even, zero--momentum, neutral interpolating operators
\begin{gather}
   P^0(t) = \frac{1}{2L^3} \sum_{\vec{x}} \{ \bar{s} \gamma_5 d(t,\vec{x}) + \bar{d} \gamma_5 s(t,\vec{x}) \}
   \ , 
   \\
   V^0_k(t) = \frac{1}{2L^3} \sum_{\vec{x}} \{ \bar{s} \gamma_k d(t,\vec{x}) - \bar{d} \gamma_k s(t,\vec{x}) \}
   \ ,
\end{gather}
and the following C--even, zero--momentum, charged interpolating operators
\begin{gather}
   P^{\{\text{s},\text{c}\}}(t) = \frac{1}{2L^3} \sum_{\vec{x}} \{ \bar{S}^{\{\text{s},\text{c}\}} \gamma_5 U^{\{\text{s},\text{c}\}}(t,\vec{x}) + \bar{U}^{\{\text{s},\text{c}\}} \gamma_5 S^{\{\text{s},\text{c}\}}(t,\vec{x}) \}
   \ , \\
   V^{\{\text{s},\text{c}\}}_k(t) = \frac{1}{2L^3} \sum_{\vec{x}} \{ \bar{S}^{\{\text{s},\text{c}\}} \gamma_k U^{\{\text{s},\text{c}\}}(t,\vec{x}) - \bar{U}^{\{\text{s},\text{c}\}} \gamma_k S^{\{\text{s},\text{c}\}}(t,\vec{x}) \}
   \ ,
\end{gather}
where the non--local operators $\bar{S}^{I}$ and $U^{I}$ are constructed as in eqs.~\eqref{eq:psistring-1}, string ($I=\text{s}$), and \eqref{eq:psicoulomb}, Coulomb ($I=\text{c}$). Under rotations the $P$ and $V$ operators transform like pseudoscalars and vectors respectively. We have calculated the following correlators
\begin{gather}
   C_{P}^I(t) = \langle P^I(t) P^I(0) \rangle \ , \qquad
   C_{V}^I(t) = \frac{1}{3} \sum_{k=1}^3 \langle V_k^I(t) V_k^I(0) \rangle \ ,
   \qquad
   I=\{0,\text{s},\text{c}\} \ .
\end{gather}
For each correlator we 
have calculated the effective mass, defined as
\begin{gather}
   M_{J}^I(t) = \cosh^{-1} \frac{C_{J}^I(t+1) + C_{J}^I(t-1)}{2 C_{J}^I(t)} \ ,
   \qquad
   J=\{P,V\}\ .
\end{gather}

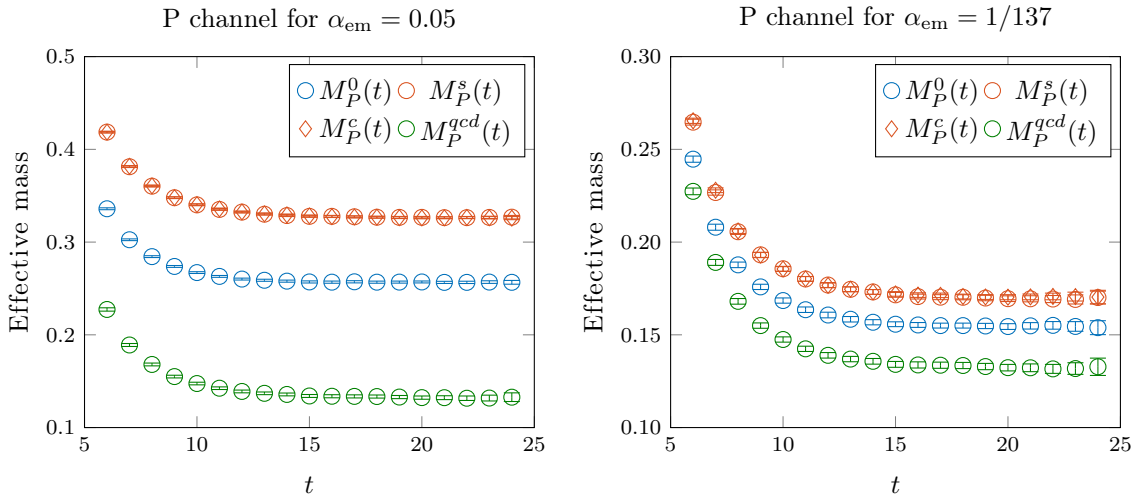
\begin{figure}[t]

\begin{tikzpicture}
\definecolor{c1}{RGB}{0,112,186}
\definecolor{c2}{RGB}{218,83,32}
\definecolor{c3}{RGB}{0,128,0}
\begin{axis}[xmin = 5, xmax = 25, ymin = 0.1, ymax = 0.5, width = 7.5cm, height = 6.5cm, xlabel = $t$, ylabel = {\textls[50]{Effective mass}}, ylabel near ticks, ticklabel style = {font=\footnotesize}, title = {P channel for $\alpha_\text{em}=0.05$}, legend style={inner sep=3pt}, legend columns=2]
\addplot[color = c1, only marks, mark = o, mark size = 3pt]
 plot [error bars/.cd, y dir = both, y explicit]
 table[x = a, y = b, y error = c] {
   a            b            c
   6.0000e+00   3.3594e-01   1.0224e-03
   7.0000e+00   3.0252e-01   9.3289e-04
   8.0000e+00   2.8432e-01   9.9184e-04
   9.0000e+00   2.7373e-01   9.9941e-04
   1.0000e+01   2.6712e-01   9.6547e-04
   1.1000e+01   2.6300e-01   9.9611e-04
   1.2000e+01   2.6012e-01   1.0343e-03
   1.3000e+01   2.5884e-01   1.1066e-03
   1.4000e+01   2.5789e-01   1.1259e-03
   1.5000e+01   2.5703e-01   1.1376e-03
   1.6000e+01   2.5680e-01   1.1792e-03
   1.7000e+01   2.5719e-01   1.1782e-03
   1.8000e+01   2.5678e-01   1.1398e-03
   1.9000e+01   2.5685e-01   1.1044e-03
   2.0000e+01   2.5701e-01   1.0496e-03
   2.1000e+01   2.5655e-01   1.1550e-03
   2.2000e+01   2.5657e-01   1.2314e-03
   2.3000e+01   2.5688e-01   1.3998e-03
   2.4000e+01   2.5663e-01   2.0353e-03
};
\addlegendentry{$M_P^0(t)$};
\addplot[color = c2, only marks, mark = o, mark size = 3pt]
 plot [error bars/.cd, y dir = both, y explicit]
 table[x = a, y = b, y error = c] {
   a            b            c
   6.0000e+00   4.1861e-01   8.5634e-04
   7.0000e+00   3.8126e-01   8.7979e-04
   8.0000e+00   3.6031e-01   8.9039e-04
   9.0000e+00   3.4784e-01   9.3546e-04
   1.0000e+01   3.4014e-01   9.3354e-04
   1.1000e+01   3.3523e-01   9.9824e-04
   1.2000e+01   3.3234e-01   8.5005e-04
   1.3000e+01   3.3009e-01   9.5987e-04
   1.4000e+01   3.2861e-01   1.0647e-03
   1.5000e+01   3.2777e-01   1.0626e-03
   1.6000e+01   3.2778e-01   1.0343e-03
   1.7000e+01   3.2729e-01   1.0309e-03
   1.8000e+01   3.2683e-01   9.9627e-04
   1.9000e+01   3.2666e-01   1.0094e-03
   2.0000e+01   3.2668e-01   1.0072e-03
   2.1000e+01   3.2634e-01   9.9363e-04
   2.2000e+01   3.2635e-01   1.0767e-03
   2.3000e+01   3.2663e-01   1.2666e-03
   2.4000e+01   3.2687e-01   1.8872e-03
};
\addlegendentry{$M_P^s(t)$};
\addplot[color = c2, only marks, mark = diamond, mark size = 3pt]
 plot [error bars/.cd, y dir = both, y explicit]
 table[x = a, y = b, y error = c] {
   a            b            c
   6.0000e+00   4.1856e-01   8.2331e-04
   7.0000e+00   3.8172e-01   8.4380e-04
   8.0000e+00   3.6063e-01   8.3699e-04
   9.0000e+00   3.4794e-01   8.7064e-04
   1.0000e+01   3.4043e-01   8.6890e-04
   1.1000e+01   3.3576e-01   9.2834e-04
   1.2000e+01   3.3252e-01   9.1853e-04
   1.3000e+01   3.3044e-01   9.8271e-04
   1.4000e+01   3.2935e-01   9.4304e-04
   1.5000e+01   3.2822e-01   9.3890e-04
   1.6000e+01   3.2747e-01   9.7673e-04
   1.7000e+01   3.2711e-01   9.9377e-04
   1.8000e+01   3.2686e-01   9.6693e-04
   1.9000e+01   3.2635e-01   9.7278e-04
   2.0000e+01   3.2609e-01   9.7496e-04
   2.1000e+01   3.2623e-01   9.7143e-04
   2.2000e+01   3.2655e-01   1.0724e-03
   2.3000e+01   3.2625e-01   1.2264e-03
   2.4000e+01   3.2585e-01   1.6735e-03
};
\addlegendentry{$M_P^c(t)$};
\addplot[color = c3, only marks, mark = o, mark size = 3pt]
 plot [error bars/.cd, y dir = both, y explicit]
 table[x = a, y = b, y error = c] {
   a            b            c
   6.0000e+00   2.2735e-01   1.8021e-03
   7.0000e+00   1.8904e-01   1.5983e-03
   8.0000e+00   1.6805e-01   1.5200e-03
   9.0000e+00   1.5498e-01   1.4478e-03
   1.0000e+01   1.4748e-01   1.4170e-03
   1.1000e+01   1.4245e-01   1.4189e-03
   1.2000e+01   1.3895e-01   1.4860e-03
   1.3000e+01   1.3693e-01   1.4421e-03
   1.4000e+01   1.3572e-01   1.5272e-03
   1.5000e+01   1.3407e-01   1.5793e-03
   1.6000e+01   1.3377e-01   1.6622e-03
   1.7000e+01   1.3362e-01   1.6668e-03
   1.8000e+01   1.3346e-01   1.6661e-03
   1.9000e+01   1.3290e-01   1.6920e-03
   2.0000e+01   1.3230e-01   1.7553e-03
   2.1000e+01   1.3223e-01   2.0892e-03
   2.2000e+01   1.3164e-01   2.4156e-03
   2.3000e+01   1.3182e-01   3.1377e-03
   2.4000e+01   1.3277e-01   4.6695e-03
};
\addlegendentry{$M_P^{qcd}(t)$};
\end{axis}
\end{tikzpicture}
\hfill
\begin{tikzpicture}
\definecolor{c1}{RGB}{0,112,186}
\definecolor{c2}{RGB}{218,83,32}
\definecolor{c3}{RGB}{0,128,0}
\begin{axis}[xmin = 5, xmax = 25, ymin = 0.1, ymax = 0.3, width = 7.5cm, height = 6.5cm, xlabel = $t$, ylabel = {\textls[50]{Effective mass}}, ylabel near ticks, ticklabel style = {font=\footnotesize}, title = {P channel for $\alpha_\text{em}=1/137$}, legend style={inner sep=3pt},y tick label style={/pgf/number format/.cd,fixed,fixed zerofill,precision=2,/tikz/.cd}, legend columns=2]
\addplot[color = c1, only marks, mark = o, mark size = 3pt]
 plot [error bars/.cd, y dir = both, y explicit]
 table[x = a, y = b, y error = c] {
   a            b            c
   6.0000e+00   2.4471e-01   1.6655e-03
   7.0000e+00   2.0802e-01   1.6290e-03
   8.0000e+00   1.8773e-01   1.4962e-03
   9.0000e+00   1.7585e-01   1.4215e-03
   1.0000e+01   1.6848e-01   1.4130e-03
   1.1000e+01   1.6354e-01   1.4824e-03
   1.2000e+01   1.6068e-01   1.5182e-03
   1.3000e+01   1.5840e-01   1.4733e-03
   1.4000e+01   1.5678e-01   1.3984e-03
   1.5000e+01   1.5572e-01   1.3001e-03
   1.6000e+01   1.5532e-01   1.3063e-03
   1.7000e+01   1.5498e-01   1.3201e-03
   1.8000e+01   1.5498e-01   1.3361e-03
   1.9000e+01   1.5477e-01   1.4071e-03
   2.0000e+01   1.5446e-01   1.5570e-03
   2.1000e+01   1.5485e-01   1.7346e-03
   2.2000e+01   1.5510e-01   2.0537e-03
   2.3000e+01   1.5452e-01   2.6007e-03
   2.4000e+01   1.5386e-01   3.7505e-03
};
\addlegendentry{$M_P^0(t)$};
\addplot[color = c2, only marks, mark = o, mark size = 3pt]
 plot [error bars/.cd, y dir = both, y explicit]
 table[x = a, y = b, y error = c] {
   a            b            c
   6.0000e+00   2.6466e-01   1.5758e-03
   7.0000e+00   2.2673e-01   1.4311e-03
   8.0000e+00   2.0558e-01   1.4555e-03
   9.0000e+00   1.9313e-01   1.3922e-03
   1.0000e+01   1.8562e-01   1.3023e-03
   1.1000e+01   1.8011e-01   1.2907e-03
   1.2000e+01   1.7685e-01   1.3248e-03
   1.3000e+01   1.7454e-01   1.3535e-03
   1.4000e+01   1.7313e-01   1.3021e-03
   1.5000e+01   1.7156e-01   1.3031e-03
   1.6000e+01   1.7062e-01   1.3117e-03
   1.7000e+01   1.7031e-01   1.3510e-03
   1.8000e+01   1.7025e-01   1.3641e-03
   1.9000e+01   1.6984e-01   1.4681e-03
   2.0000e+01   1.6949e-01   1.5414e-03
   2.1000e+01   1.6943e-01   1.6484e-03
   2.2000e+01   1.6930e-01   1.9070e-03
   2.3000e+01   1.6908e-01   2.4510e-03
   2.4000e+01   1.7010e-01   3.4249e-03
};
\addlegendentry{$M_P^s(t)$};
\addplot[color = c2, only marks, mark = diamond, mark size = 3pt]
 plot [error bars/.cd, y dir = both, y explicit]
 table[x = a, y = b, y error = c] {
   a            b            c
   6.0000e+00   2.6540e-01   1.4168e-03
   7.0000e+00   2.2795e-01   1.2833e-03
   8.0000e+00   2.0600e-01   1.3304e-03
   9.0000e+00   1.9300e-01   1.3776e-03
   1.0000e+01   1.8547e-01   1.2844e-03
   1.1000e+01   1.8021e-01   1.2583e-03
   1.2000e+01   1.7666e-01   1.2593e-03
   1.3000e+01   1.7459e-01   1.2500e-03
   1.4000e+01   1.7312e-01   1.2801e-03
   1.5000e+01   1.7206e-01   1.2445e-03
   1.6000e+01   1.7142e-01   1.2414e-03
   1.7000e+01   1.7129e-01   1.2240e-03
   1.8000e+01   1.7065e-01   1.2922e-03
   1.9000e+01   1.7040e-01   1.3045e-03
   2.0000e+01   1.7029e-01   1.3720e-03
   2.1000e+01   1.7045e-01   1.5545e-03
   2.2000e+01   1.7070e-01   1.7925e-03
   2.3000e+01   1.7062e-01   2.3873e-03
   2.4000e+01   1.7035e-01   3.6633e-03
};
\addlegendentry{$M_P^c(t)$};
\addplot[color = c3, only marks, mark = o, mark size = 3pt]
 plot [error bars/.cd, y dir = both, y explicit]
 table[x = a, y = b, y error = c] {
   a            b            c
   6.0000e+00   2.2735e-01   1.8021e-03
   7.0000e+00   1.8904e-01   1.5983e-03
   8.0000e+00   1.6805e-01   1.5200e-03
   9.0000e+00   1.5498e-01   1.4478e-03
   1.0000e+01   1.4748e-01   1.4170e-03
   1.1000e+01   1.4245e-01   1.4189e-03
   1.2000e+01   1.3895e-01   1.4860e-03
   1.3000e+01   1.3693e-01   1.4421e-03
   1.4000e+01   1.3572e-01   1.5272e-03
   1.5000e+01   1.3407e-01   1.5793e-03
   1.6000e+01   1.3377e-01   1.6622e-03
   1.7000e+01   1.3362e-01   1.6668e-03
   1.8000e+01   1.3346e-01   1.6661e-03
   1.9000e+01   1.3290e-01   1.6920e-03
   2.0000e+01   1.3230e-01   1.7553e-03
   2.1000e+01   1.3223e-01   2.0892e-03
   2.2000e+01   1.3164e-01   2.4156e-03
   2.3000e+01   1.3182e-01   3.1377e-03
   2.4000e+01   1.3277e-01   4.6695e-03
};
\addlegendentry{$M_P^{qcd}(t)$};
\end{axis}
\end{tikzpicture}
\caption{\label{fig:pseudosplitting} 
Effective masses for the pseudoscalar correlators at $\alpha_\text{em}=0.05$ (left plot, blue and orange points), $\alpha_\text{em}=1/137$ (right plot, blue and orange points), and $\alpha_\text{em}=0$ (both plots, green points). The blue points correspond to the neutral states, $M_{P}^0(t)$. The orange points correspond to the charged states interpolated by using either the string, $M_{P}^\text{s}(t)$, or the Coulomb, $M_{P}^\text{c}(t)$, operators. The green points correspond to the states in QCD--only simulations, $M_{P}^{qcd}(t)$. The quality of the numerical signal is essentially the same for charged and neutral states, with and without QED, and it is not affected by the non--local gauge--invariant operators used in the charged channel. With these unphysical values of the bare parameters the charged--neutral mass splitting can be extracted with statistical significance even at $\alpha_\text{em}=1/137$.}
\end{figure}

\begin{figure}[t]

\begin{tikzpicture}
\definecolor{c1}{RGB}{0,112,186}
\definecolor{c2}{RGB}{218,83,32}
\begin{axis}[xmin = 5, xmax = 25, ymin = 0.3, ymax = 0.7, width = 7.5cm, height = 5.5cm, xlabel = $t$, ylabel = {\textls[50]{Effective mass}}, ylabel near ticks, ticklabel style = {font=\footnotesize}, title = {V channel for $\alpha_\text{em}=0.05$}, legend style={inner sep=3pt}]
\addplot[color = c1, only marks, mark = o, mark size = 3pt]
 plot [error bars/.cd, y dir = both, y explicit]
 table[x = a, y = b, y error = c] {
   a            b            c
   6.0000e+00   5.9517e-01   7.2209e-04
   7.0000e+00   5.2524e-01   8.2072e-04
   8.0000e+00   4.8229e-01   9.5697e-04
   9.0000e+00   4.5651e-01   1.1494e-03
   1.0000e+01   4.3886e-01   1.3442e-03
   1.1000e+01   4.2635e-01   1.5960e-03
   1.2000e+01   4.1828e-01   1.7894e-03
   1.3000e+01   4.1344e-01   1.9301e-03
   1.4000e+01   4.0934e-01   2.3373e-03
   1.5000e+01   4.0597e-01   2.6968e-03
   1.6000e+01   4.0576e-01   3.0232e-03
   1.7000e+01   4.0593e-01   3.6746e-03
   1.8000e+01   4.0271e-01   4.4057e-03
   1.9000e+01   4.0047e-01   5.1181e-03
   2.0000e+01   3.9966e-01   5.9057e-03
   2.1000e+01   3.9367e-01   6.9044e-03
   2.2000e+01   3.9781e-01   8.1356e-03
   2.3000e+01   3.9692e-01   1.0579e-02
   2.4000e+01   3.8005e-01   1.7231e-02
};
\addlegendentry{$M_V^0(t)$};
\addplot[color = c2, only marks, mark = o, mark size = 3pt]
 plot [error bars/.cd, y dir = both, y explicit]
 table[x = a, y = b, y error = c] {
   a            b            c
   6.0000e+00   6.2807e-01   6.4008e-04
   7.0000e+00   5.6310e-01   6.9440e-04
   8.0000e+00   5.2337e-01   8.2091e-04
   9.0000e+00   4.9772e-01   8.6450e-04
   1.0000e+01   4.8103e-01   9.2077e-04
   1.1000e+01   4.6896e-01   1.0855e-03
   1.2000e+01   4.6322e-01   1.2803e-03
   1.3000e+01   4.5800e-01   1.3938e-03
   1.4000e+01   4.5275e-01   1.5856e-03
   1.5000e+01   4.5066e-01   1.7368e-03
   1.6000e+01   4.4925e-01   2.0271e-03
   1.7000e+01   4.4595e-01   2.1846e-03
   1.8000e+01   4.4503e-01   2.7496e-03
   1.9000e+01   4.4145e-01   2.8136e-03
   2.0000e+01   4.4418e-01   3.1700e-03
   2.1000e+01   4.4256e-01   3.7916e-03
   2.2000e+01   4.4429e-01   4.6481e-03
   2.3000e+01   4.3918e-01   5.7256e-03
   2.4000e+01   4.2729e-01   9.2522e-03
};
\addlegendentry{$M_V^s(t)$};
\addplot[color = c2, only marks, mark = diamond, mark size = 3pt]
 plot [error bars/.cd, y dir = both, y explicit]
 table[x = a, y = b, y error = c] {
   a            b            c
   6.0000e+00   6.2776e-01   6.2477e-04
   7.0000e+00   5.6312e-01   6.3850e-04
   8.0000e+00   5.2355e-01   7.3843e-04
   9.0000e+00   4.9854e-01   7.9220e-04
   1.0000e+01   4.8175e-01   8.9741e-04
   1.1000e+01   4.7046e-01   9.8707e-04
   1.2000e+01   4.6317e-01   1.1207e-03
   1.3000e+01   4.5811e-01   1.2688e-03
   1.4000e+01   4.5412e-01   1.4053e-03
   1.5000e+01   4.5021e-01   1.6857e-03
   1.6000e+01   4.4629e-01   1.8330e-03
   1.7000e+01   4.4493e-01   1.9886e-03
   1.8000e+01   4.4127e-01   2.3323e-03
   1.9000e+01   4.4197e-01   2.7653e-03
   2.0000e+01   4.4108e-01   3.3068e-03
   2.1000e+01   4.3717e-01   3.5417e-03
   2.2000e+01   4.4243e-01   4.4667e-03
   2.3000e+01   4.4513e-01   5.4403e-03
   2.4000e+01   4.3725e-01   7.9196e-03
};
\addlegendentry{$M_V^c(t)$};
\end{axis}
\end{tikzpicture}
\hfill
\begin{tikzpicture}
\definecolor{c1}{RGB}{0,112,186}
\definecolor{c2}{RGB}{218,83,32}
\begin{axis}[xmin = 5, xmax = 25, ymin = 0.2, ymax = 0.6, width = 7.5cm, height = 5.5cm, xlabel = $t$, ylabel = {\textls[50]{Effective mass}}, ylabel near ticks, ticklabel style = {font=\footnotesize}, title = {V channel for $\alpha_\text{em}=1/137$}, legend style={inner sep=3pt}]
\addplot[color = c1, only marks, mark = o, mark size = 3pt]
 plot [error bars/.cd, y dir = both, y explicit]
 table[x = a, y = b, y error = c] {
   a            b            c
   6.0000e+00   5.6228e-01   1.2056e-03
   7.0000e+00   4.7964e-01   1.4211e-03
   8.0000e+00   4.2745e-01   1.6990e-03
   9.0000e+00   3.9163e-01   1.9244e-03
   1.0000e+01   3.6707e-01   2.2867e-03
   1.1000e+01   3.4969e-01   2.5983e-03
   1.2000e+01   3.3611e-01   3.1687e-03
   1.3000e+01   3.2557e-01   3.5270e-03
   1.4000e+01   3.1490e-01   3.9219e-03
   1.5000e+01   3.0138e-01   4.7417e-03
   1.6000e+01   2.9740e-01   5.2568e-03
   1.7000e+01   2.9193e-01   6.1265e-03
   1.8000e+01   2.9574e-01   7.6474e-03
   1.9000e+01   2.9302e-01   9.1846e-03
   2.0000e+01   2.9284e-01   1.1172e-02
   2.1000e+01   2.9426e-01   1.3283e-02
   2.2000e+01   2.9105e-01   1.6226e-02
   2.3000e+01   2.8848e-01   2.0272e-02
   2.4000e+01   3.0958e-01   3.1894e-02
};
\addlegendentry{$M_V^0(t)$};
\addplot[color = c2, only marks, mark = o, mark size = 3pt]
 plot [error bars/.cd, y dir = both, y explicit]
 table[x = a, y = b, y error = c] {
   a            b            c
   6.0000e+00   5.6456e-01   1.0323e-03
   7.0000e+00   4.8394e-01   1.1775e-03
   8.0000e+00   4.3246e-01   1.4195e-03
   9.0000e+00   3.9844e-01   1.7537e-03
   1.0000e+01   3.7415e-01   1.9132e-03
   1.1000e+01   3.5663e-01   2.2291e-03
   1.2000e+01   3.4467e-01   2.6162e-03
   1.3000e+01   3.3402e-01   3.0764e-03
   1.4000e+01   3.2914e-01   3.4959e-03
   1.5000e+01   3.2618e-01   4.2494e-03
   1.6000e+01   3.1667e-01   4.7754e-03
   1.7000e+01   3.1749e-01   5.5328e-03
   1.8000e+01   3.1728e-01   6.4280e-03
   1.9000e+01   3.1645e-01   7.8231e-03
   2.0000e+01   3.1386e-01   9.0796e-03
   2.1000e+01   3.2123e-01   1.1639e-02
   2.2000e+01   3.1736e-01   1.4142e-02
   2.3000e+01   3.3786e-01   1.8678e-02
   2.4000e+01   3.4707e-01   2.9065e-02
};
\addlegendentry{$M_V^s(t)$};
\addplot[color = c2, only marks, mark = diamond, mark size = 3pt]
 plot [error bars/.cd, y dir = both, y explicit]
 table[x = a, y = b, y error = c] {
   a            b            c
   6.0000e+00   5.6518e-01   1.0783e-03
   7.0000e+00   4.8349e-01   1.2579e-03
   8.0000e+00   4.3246e-01   1.4873e-03
   9.0000e+00   4.0002e-01   1.7063e-03
   1.0000e+01   3.7711e-01   1.9840e-03
   1.1000e+01   3.5924e-01   2.3930e-03
   1.2000e+01   3.4543e-01   2.7605e-03
   1.3000e+01   3.3309e-01   3.1900e-03
   1.4000e+01   3.2614e-01   3.7911e-03
   1.5000e+01   3.1763e-01   4.1146e-03
   1.6000e+01   3.1267e-01   5.0546e-03
   1.7000e+01   3.1084e-01   5.6183e-03
   1.8000e+01   3.0705e-01   6.5894e-03
   1.9000e+01   2.9341e-01   7.2679e-03
   2.0000e+01   3.0814e-01   8.9370e-03
   2.1000e+01   3.0916e-01   1.1276e-02
   2.2000e+01   3.0953e-01   1.3240e-02
   2.3000e+01   2.6670e-01   1.7394e-02
   2.4000e+01   2.7682e-01   2.8300e-02
};
\addlegendentry{$M_V^c(t)$};
\end{axis}
\end{tikzpicture}
\caption{\label{fig:vectorsplitting} 
Effective masses for the vector correlators at $\alpha_\text{em}=0.05$ (left plot) and $\alpha_\text{em}=1/137$ (right plot). In both plots the blue points correspond to the neutral states, $M_{V}^0(t)$, while the orange points correspond to the charged states interpolated by using either the string, $M_{V}^\text{s}(t)$, or the Coulomb, $M_{V}^\text{c}(t)$, operators. Also in this channel the quality of the numerical signal is essentially the same for charged and neutral states. With these unphysical values of the bare parameters the charged--neutral mass splitting can be extracted with statistical significance at $\alpha_\text{em}=0.05$ while more statistics is required at $\alpha_\text{em}=1/137$. This is not surprising as vector correlators have a worse signal--to--noise ratio with respect to pseudoscalar ones.}
\end{figure}
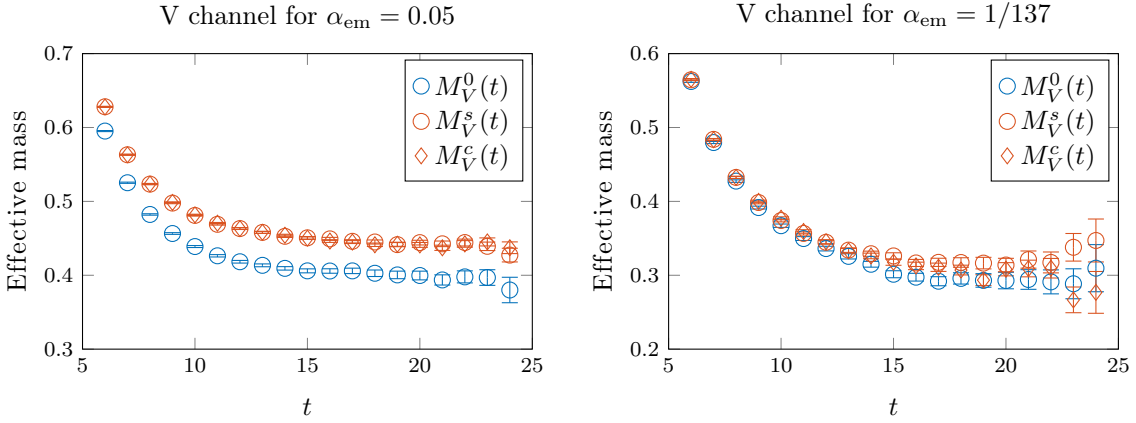

The effective masses are shown in fig.~\ref{fig:pseudosplitting} for the P states and in fig.~\ref{fig:vectorsplitting} for the V states for both values of $\alpha_\text{em}$. For comparison, in fig.~\ref{fig:pseudosplitting} we also report the effective mass calculated on a QCD--only ensemble, generated with the CLS H200 bare parameters on a $48 \times 24^3$ lattice with C$^\star$ boundary conditions. In all cases we have used 500 configurations and 8 stochastic sources per configuration. As expected, we observe that the pseudoscalar masses are larger with respect to the ones quoted in ref.~\cite{Bruno:2014jqa} for the H200 ensemble, because of the mass shift due to the electromagnetic interactions. 

The most important result of this paper is the fact that effective masses can be extracted with similar errors in the neutral and charged channels. In fact, the introduction of the non--local gauge--invariant operators for charged states does not affect much the quality of the signal in correlators and effective masses. We also observe that in these channels, the string and Coulomb operators behave very similarly. Moreover, we observe that the statistical errors in the QCD+QED pseudoscalar effective mass are very similar to their QCD--only counterparts.

While these simulations are performed at unphysical values of the quark masses, the charged--neutral mass splittings can clearly be extracted with a statistically significant accuracy for both the pseudoscalar and vector states at $\alpha_\text{em}=0.05$. Remarkably, the mass splitting in the pseudoscalar channel is statistically significant even at $\alpha_\text{em}=1/137$.

\subsection{Charged mesons with real photons}
\label{sec:numericsB}
The goal of this subsection is to sketch a strategy to extract states of charged mesons with real photons. Let us focus on the charged vector channel. In finite volume, the spectral decomposition can be written for the V correlator. Amplitudes can be organised according to their leading behaviour in $\alpha_\text{em}$, i.e.
\begin{gather}
   C_{V}^I(t) = \sum_{n=0}^\infty \sum_{r=0}^\infty c^I_{n,r} e^{-E_{n,r}t} \ , 
   \nonumber \\
   c_{n,r}^I = \frac{L^3}{3} \sum_{k=1}^3 \bigg| \langle \Omega | \tfrac{1}{L^3} \sum_{\vec{x}} V_k^I(\vec{x}) | n,r \rangle \bigg|^2 = O(\alpha_\text{em}^r) \ ,
   \qquad
   I=\{\text{s},\text{c} \}\;.
\end{gather}
In these formulae we assume that the $T \to \infty$ limit has been taken already. Since the full QCD+QED Hamiltonian does not conserve the photon number, the states $|n,r\rangle$ are not eigenstates of the photon number operator. However at the leading order in $\alpha_\text{em}^{1/2}$, the state $|n,r\rangle$ is nothing but the tensor product of a QCD state with $r$ free real photons, and its energy is given by the energy of the QCD state plus the energy of the free photons. Therefore it makes sense to refer to $|n,r\rangle$ as a state with $r$ real photons, as long as $\alpha_\text{em}$ is small enough. Notice that these states are gauge invariant by construction, therefore only physical polarizations of the photon contribute.

If the volume is large enough, the ground state of the $C_V^I(t)$ correlator is a state with one real photon. At the leading order in $\alpha_\text{em}^{1/2}$ this state contains a charged P particle and a real photon in a kinematic configuration with zero momentum and zero angular momentum. Some tedious but standard group theory reveals that, in order to be able to construct a state in the vector ($\text{T}_1^-$) representation of the cubic group $O_h$, the minimum--norm momentum allowed for the photon is
\begin{gather}
   \bar{\vec{p}} = \frac{\pi}{L} (1,1,1) \ ,
\end{gather}
up to isometries of the cube\footnote{We remind that, because of C$^\star$ boundary conditions, the photon field is antiperiodic in all spatial directions. Therefore the allowed momenta for the photons have components that are odd multiples of $\pi/L$.}. This state has energy equal to
\begin{gather}
   E_{0,1} = \sqrt{ M_P^2 + \frac{3 \pi^2}{L^2} } + \frac{\pi \sqrt{3}}{L} + O(\alpha_\text{em}) \ ,
\end{gather}
and is created at the leading order in $\alpha_\text{em}^{1/2}$ by the following interpolating operator
\begin{gather}
   W^I_k(t) = \sum_{\vec{p} \in O_h \bar{\vec{p}}} \tilde{P}^I(t,-\vec{p}) \epsilon_{k\ell j} p_\ell \tilde{A}^\text{c}_j(t,\vec{p}) \ ,
\end{gather}
where $\tilde{P}^I$ and $\tilde{A}^\text{c}$ are defined as
\begin{gather}
   \tilde{P}^I(t,\vec{p}) = \frac{1}{2L^3} \sum_{\vec{x}} e^{i\vec{p}\vec{x}} \{ \bar{S}^I \gamma_5 U^I(t,\vec{x}) - \bar{U}^I \gamma_5 S^I(t,\vec{x}) \}
   \ , 
   \\
   \tilde{A}^{\text{c}}_k(t,\vec{p}) = \frac{1}{L^3} \sum_{\vec{x}} e^{i\vec{p}\vec{x}} A^\text{c}_k(t,\vec{x})
   \ ,
\end{gather}
and $A^\text{c}_k$ is the gauge--invariant representation of the Coulomb--gauge photon field defined in eq.~\eqref{eq:Ac}. Notice that the operator $\tilde{P}^I(t,\vec{p})$ is C--odd, contrarily to the analogous operator defined in the previous subsection. This is due to the fact that states with momentum $\bar{\vec{p}}$ are antiperiodic, i.e. they are odd under a translation by a distance $L$ in any of the spatial directions, and therefore odd under charge conjugation.

If the volume is large enough and $\alpha_\text{em}$ is small enough, then the inequality $E_{0,1} < E_{0,0}$ comes from the observation that $M_P$ is always smaller than $M_V$, and in particular this is true at $\alpha_\text{em}=0$. However as the volume goes to zero, the relative momentum of the two particles in the $|0,1\rangle$ state diverge and so does $E_{0,1}$. Therefore, if the volume is small enough, then $E_{0,1} > E_{0,0}$. It will turn out that this is the kinematic region of our simulations.

\begin{figure}[!t]
\begin{tikzpicture}
\definecolor{c1}{RGB}{0,112,186}
\definecolor{c2}{RGB}{218,83,32}
\begin{axis}[xmin = 5, xmax = 25, ymin = 0.3, ymax = 0.9, width = 7.5cm, height = 5.5cm, xlabel = $t$, ylabel = {\textls[50]{Effective mass}}, ylabel near ticks, ticklabel style = {font=\footnotesize}, title = {V channel for $\alpha_\text{em}=0.05$}, legend style={inner sep=3pt}]
\addplot[color = c1, only marks, mark = o, mark size = 3pt]
 plot [error bars/.cd, y dir = both, y explicit]
 table[x = a, y = b, y error = c] {
   a            b            c
   6.0000e+00   6.2801e-01   1.0987e-03
   7.0000e+00   5.6327e-01   1.2889e-03
   8.0000e+00   5.2417e-01   1.4221e-03
   9.0000e+00   4.9941e-01   1.5923e-03
   1.0000e+01   4.8237e-01   1.7682e-03
   1.1000e+01   4.7126e-01   2.0313e-03
   1.2000e+01   4.6485e-01   2.5281e-03
   1.3000e+01   4.6115e-01   2.5780e-03
   1.4000e+01   4.5550e-01   2.9308e-03
   1.5000e+01   4.5195e-01   3.3622e-03
   1.6000e+01   4.5040e-01   3.9863e-03
   1.7000e+01   4.4646e-01   4.5667e-03
   1.8000e+01   4.4373e-01   4.9131e-03
   1.9000e+01   4.3956e-01   5.7246e-03
   2.0000e+01   4.4179e-01   6.1458e-03
   2.1000e+01   4.3918e-01   7.4471e-03
   2.2000e+01   4.4362e-01   9.2702e-03
   2.3000e+01   4.4422e-01   1.1117e-02
   2.4000e+01   4.2772e-01   1.8497e-02
};
\addlegendentry{$\lambda_0(t,8)$};
\addplot[color = c2, only marks, mark = o, mark size = 3pt]
 plot [error bars/.cd, y dir = both, y explicit]
 table[x = a, y = b, y error = c] {
   a            b            c
   6.0000e+00   7.4480e-01   1.4869e-02
   7.0000e+00   7.0258e-01   1.7116e-02
   8.0000e+00   6.6607e-01   2.0690e-02
   9.0000e+00   6.4797e-01   2.9622e-02
   1.0000e+01   6.3222e-01   3.7692e-02
   1.1000e+01   6.5933e-01   4.8831e-02
   1.2000e+01   5.8615e-01   5.9990e-02
   1.3000e+01   5.7349e-01   7.5221e-02
   1.4000e+01   6.3653e-01   9.8310e-02
   1.5000e+01   5.8373e-01   1.0530e-01
};
\addlegendentry{$\lambda_1(t,8)$};
\end{axis}
\end{tikzpicture}
\hfill
\begin{tikzpicture}
\definecolor{c1}{RGB}{0,112,186}
\definecolor{c2}{RGB}{218,83,32}
\begin{axis}[xmin = 5, xmax = 25, ymin = 0.2, ymax = 0.8, width = 7.5cm, height = 5.5cm, xlabel = $t$, ylabel = {\textls[50]{Effective mass}}, ylabel near ticks, ticklabel style = {font=\footnotesize}, title = {V channel for $\alpha_\text{em}=1/137$}, legend style={inner sep=3pt}]
\addplot[color = c1, only marks, mark = o, mark size = 3pt]
 plot [error bars/.cd, y dir = both, y explicit]
 table[x = a, y = b, y error = c] {
   a            b            c
   6.0000e+00   5.6519e-01   1.8928e-03
   7.0000e+00   4.8412e-01   2.2379e-03
   8.0000e+00   4.3244e-01   2.7295e-03
   9.0000e+00   3.9944e-01   3.3555e-03
   1.0000e+01   3.7670e-01   3.9929e-03
   1.1000e+01   3.5878e-01   4.4968e-03
   1.2000e+01   3.4780e-01   5.3211e-03
   1.3000e+01   3.3524e-01   6.0876e-03
   1.4000e+01   3.3018e-01   7.2629e-03
   1.5000e+01   3.1970e-01   8.5284e-03
   1.6000e+01   3.0926e-01   9.2081e-03
   1.7000e+01   3.1029e-01   1.0448e-02
   1.8000e+01   3.0616e-01   1.2542e-02
   1.9000e+01   3.0327e-01   1.5164e-02
   2.0000e+01   3.1279e-01   1.9087e-02
   2.1000e+01   3.2059e-01   2.3891e-02
   2.2000e+01   3.2444e-01   3.2036e-02
   2.3000e+01   3.1667e-01   4.1934e-02
   2.4000e+01   3.3091e-01   7.5018e-02
};
\addlegendentry{$\lambda_0(t,8)$};
\addplot[color = c2, only marks, mark = o, mark size = 3pt]
 plot [error bars/.cd, y dir = both, y explicit]
 table[x = a, y = b, y error = c] {
   a            b            c
   6.0000e+00   6.6740e-01   1.4410e-02
   7.0000e+00   6.0706e-01   1.8749e-02
   8.0000e+00   5.9484e-01   2.6244e-02
   9.0000e+00   5.4682e-01   3.1987e-02
   1.0000e+01   5.8461e-01   4.5908e-02
   1.1000e+01   4.9138e-01   5.8721e-02
   1.2000e+01   5.2088e-01   7.2269e-02
   1.3000e+01   4.3104e-01   9.3072e-02
   1.4000e+01   5.7583e-01   1.5222e-01
};
\addlegendentry{$\lambda_1(t,8)$};
\end{axis}
\end{tikzpicture}
\caption{\label{fig:gevp} 
Effective masses for the ground state, $\lambda_0(t,8)$ (blue points), and the excited state, $\lambda_1(t,8)$ (orange points), eigenvalues obtained by solving eq.~(\ref{eq:gevp}) for both the string and Coulomb operators and by averaging the corresponding results. The left plot corresponds to $\alpha_\text{em}=0.05$ while the right plot to $\alpha_\text{em}=1/137$.}
\end{figure}
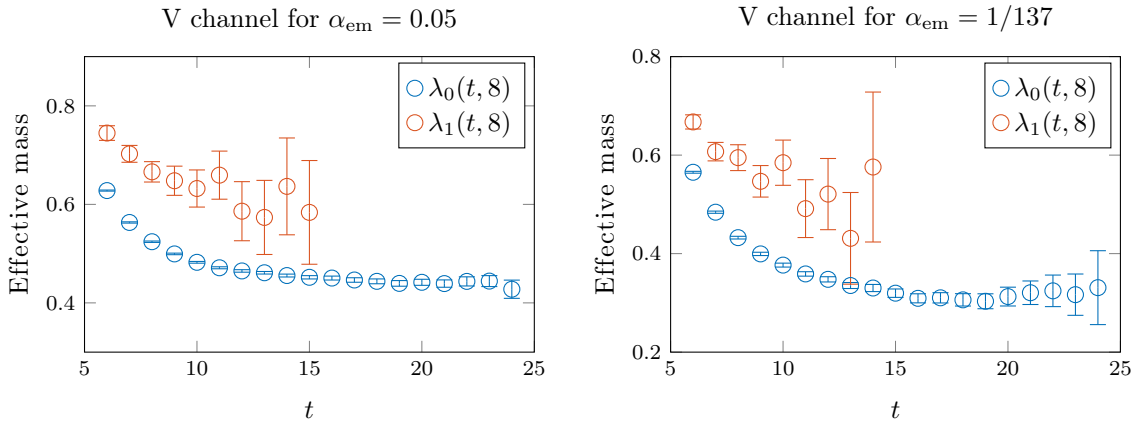

One can set up a generalised eigenvalue problem with two operators: $V_k^I$ and $W^I_k$. If $\alpha_\text{em}$ is small enough, $V_k^I$ has maximal overlap with the state $|0,0\rangle$ and $W^I_k$ has maximal overlap with the state $|0,1\rangle$. At moderate value of $\alpha_\text{em}$, or in the regime in which the P$+\gamma$ state is almost degenerate with a P$+$P state (which is in fact the case in our simulations), a larger operator basis may be necessary. In this exploratory calculation we will ignore these subtleties and proceed with the simple two--operator setup. If $C^I(t)$ is the $2 \times 2$ matrix of correlators constructed with the operators $V_k^I$ and $W^I_k$, we solve the generalised eigenvalue problem given by
\begin{gather}
   C^I(t)\, v_n^I(t,t_0) = \lambda_n^I(t,t_0)\, C^I(t_0)\, v_n^I(t,t_0) \ , \qquad n=0,1.
   \label{eq:gevp}
\end{gather}
We have extracted the ground state, $\lambda_0^I(t,t_0)$, and the excited state, $\lambda_1^I(t,t_0)$, eigenvalues by using both the string and Coulomb interpolating operators by obtaining statistically consistent results with essentially the same quality of the signal--to--noise ratio.  
In fig.~\ref{fig:gevp} we plot the effective masses extracted from 
\begin{gather}
\lambda_n(t,t_0)=\frac{\lambda_n^\text{s}(t,t_0)+\lambda_n^\text{c}(t,t_0)}{2}
\end{gather}
for $n=0,1$, corresponding to $\alpha_\text{em}=1/137$ and $\alpha_\text{em}=0.05$ respectively. The presented results are obtained with $t_0=8$, but we have checked the stability of our results in the range $t_0 \in [4,10]$.

On the one hand, from a quantitative analysis of the excited--state energy it turns out that (as anticipated in the discussion above) we cannot discriminate between a P$+\gamma$ and a P$+$P state within the present statistical uncertainties. Since this may be due to the unphysical values of the bare parameters used in this study, we postpone a more detailed numerical analysis to future work on this subject. This will certainly require more statistics and, possibly, an extended basis of interpolating operators.

On the other hand, some qualitative information can be drawn from the plots in fig.~\ref{fig:gevp}. In our opinion, the quality of the numerical signals makes us pretty confident of the possibility to probe charged states containing real photons by using a fully non--perturbative gauge--invariant strategy along the lines of the one sketched in this section.

\section{Conclusions}
\label{sec:conclusions}

We have performed numerical lattice simulations of the compact formulation of QCD$+$QED with C--periodic boundary conditions in the spatial directions. In this setup, following ref.~\cite{Lucini:2015hfa}, charged--hadron masses can be calculated from first principles without relying on gauge fixing at any stage of the calculation. 

Our simulations are performed at unphysical values of the bare parameters, with pseudoscalar meson masses of the order of the physical kaon at $\alpha_\text{em}=1/137$. For this reason our results do not have phenomenological relevance but do have, in our opinion, deep theoretical implications. We provide a clear evidence that the strategy of ref.~\cite{Lucini:2015hfa} is numerically viable and that charged states can be efficiently probed in a gauge--invariant way.

In particular, we show in section~\ref{sec:numerics} that the masses of charged hadrons can be extracted with the same numerical accuracy as their almost--degenerate neutral counterparts. This is true both in the pseudoscalar and in the vector meson channels. At the values of the bare parameters used in our study, the pseudoscalar--meson charged--neutral mass splitting can be extracted with statistical significance even in the simulation performed at $\alpha_\text{em}=1/137$.  

We have also sketched a strategy to probe states of charged mesons with real photons. The proposal consists of using gauge--invariant interpolating operators that, at leading order in $\alpha_\text{em}$, have maximal overlap with states having a fixed number of real photons. Although much more work is certainly needed in this direction, the results of subsection~\ref{sec:numericsB} represent a promising indication on the numerical validity of this approach.

\begin{acknowledgments}
This work is part of the programme of the RC$^\star$ Collaboration and we warmly thank our colleagues for their help. We are particularly indebted to Alberto Ramos for his contribution to various stages of this work. BL is supported in part by the Royal Society, by the Wolfson Foundation and by the STFC Consolidated Grants ST/L000369/1 and ST/P00055X/1. MH is supported by the Danish National Research Foundation grant DNRF90 and by a Lundbeck Foundation Fellowship grant. Numerical simulations have been performed on clusters of the Supercomputing Wales project, partly funded by the European Regional Development Fund (ERDF) via Welsh Government, on a cluster at CERN, managed by the HPC team in the IT Department, and on the Marconi system at CINECA under the initiative INFN-LQCD123.
\end{acknowledgments}


\appendix

\section{Gauge--fixed two--point functions}
\label{sec:gftpf}

The goal of this appendix is to illustrate some of the subtleties that arise in the charged sector, when the U(1) gauge is fixed. For definiteness we work here with the familiar case of covariant gauge, in continuum notation. In order to avoid potential issues with IR divergences, we consider QCD$+$QED in a spatial box with size $L^3$ and C$^\star$ boundary conditions for all fields. For simplicity we consider an infinite time extent. In Euclidean spacetime, the action in covariant gauge is
\begin{gather}
   S_{\xi_0} = S_0(A,B,\psi,\bar{\psi}) + \frac{\xi_0}{2e_0^2} ( \partial_\mu A_\mu , \partial_\nu A_\nu )
   \ ,
\end{gather}
where $S_0$ is the gauge--invariant part of the action, $A_\mu$ and $B_\mu$ are the photon and gluon fields, while $\psi$ and $\bar{\psi}$ are the quark fields, and the scalar product is defined as
\begin{gather}
   ( f , g ) = \int d^4 x \ f(x)^* g(x) \ .
\end{gather}
Let $h(x)$ be some local operator which interpolates a hadron with electric charge $q_h$, and let $\bar{h}(x)$ the interpolating operator with the corresponding antiparticle. We are interested in the two--point function
\begin{gather}
   \langle h(y) \bar{h}(x) \rangle_{\xi_0}
   =
   \frac{
   \int [d \lambda] \, [dA] \, [dB] \, [d\psi] \, [d\bar{\psi}] e^{-S_{\xi_0}(A,B,\psi,\bar{\psi})} h(y) \bar{h}(x)
   }{
   \int [d \lambda] \, [dA] \, [dB] \, [d\psi] \, [d\bar{\psi}] e^{-S_{\xi_0}(A,B,\psi,\bar{\psi})}
   }
   \ .
\end{gather}
The integrands do not depend on $\lambda$, therefore the auxiliary integral over $\lambda$ gives an infinite constant which simplifies in the ratio. We change variables in the two integrals to the gauge--transformed fields
\begin{gather}
   A_\mu(x) \to A_\mu(x) + \partial_\mu \lambda(x) \ , \qquad
   \psi_f(x) \to \exp\{\imath q_f \lambda(x)\} \psi_f(x) \ .
\end{gather}
The interpolating operator and action transform as
\begin{gather}
   h(x) \to \exp\{\imath q_h \lambda(x)\} h(x) \ , \qquad
   S_{\xi_0}  \to S_0 + \frac{\xi_0}{2e_0^2} ( \partial_\mu A_\mu + \Box \lambda , \partial_\mu A_\mu + \Box \lambda )
   \ .
\end{gather}
After this change of variables, the integral over $\lambda$ is Gaussian and can be calculated analytically, yielding the following gauge--invariant representation
\begin{gather}
   \langle h(z) \bar{h}(y) \rangle_{\xi_0}
   = 
   e^{-\frac{1}{2\xi_0} ( J_\mu , \frac{1}{-\Box}  J_\mu)}
   \langle e^{-\imath (J_\mu , A_\mu)} h(z) \bar{h}(y) \rangle_{0}
   \ , 
   \label{eq:2ptf-gi}
\end{gather}
where the current $J_\mu(z)$ is defined by the equation
\begin{gather}
   \Box J_\mu(x) = q_h \partial_\mu [ \delta^4(x-y) - \delta^4(x-z) ] \ .
\end{gather}
Because of C$^\star$ boundary conditions, the Laplacian $\Box = \partial_\mu \partial_\mu$ is defined with antiperiodic boundary conditions in space and is therefore invertible. The expectation value in eq.~\eqref{eq:2ptf-gi} is calculated with the gauge--invariant action $S_0$. Under a gauge transformation $\lambda(x)$ with antiperiodic boundary conditions in space the phase factor in eq.~\eqref{eq:2ptf-gi} transforms as
\begin{flalign}
   e^{-\imath (J_\mu , A_\mu)}
   \to & 
   e^{-\imath (J_\mu , A_\mu) - \imath (J_\mu , \partial_\mu \lambda)}
   = \nonumber \\ & =
   e^{-\imath (J_\mu , A_\mu) + \imath (\partial_\mu J_\mu , \lambda)}
   =
   e^{-\imath (J_\mu , A_\mu)} e^{\imath q_h [\lambda(y)-\lambda(z)]}
   \ .
\end{flalign}
The integration by part $(J_\mu , \partial_\mu \lambda) = - (\partial_\mu J_\mu , \lambda)$ does not generate boundary terms since the product $J_\mu \lambda$ satisfies periodic boundary conditions. The factor $e^{\imath q_h [\lambda(y)-\lambda(z)]}$ in the above equation cancels the phase generated by the gauge transformation of $h(z) \bar{h}(y)$. As a consequence, the observable in eq.~\eqref{eq:2ptf-gi} is invariant under local gauge transformations. It is tempting to interpret the gauge--invariant observable
\begin{gather}
   H(x) = e^{\imath q_h (\frac{1}{\Box} \partial_\mu  \delta_{x} , A_\mu)} h(x)~,
\end{gather}
as a possible interpolating operator for the charged hadron $h$. In fact this operator is formally very similar to Dirac's interpolating operator. However $H(x)$ is non--local in time and a standard interpretation as an interpolating operator is not possible. The Hamiltonian representation of the expectation value in the r.h.s. of eq.~\eqref{eq:2ptf-gi} is obtained by interpreting the phase as a term of the action. As in the case of $J=0$, the action $S_0 + \imath (J_\mu , A_\mu)$ defines a constrained Hamiltonian system. States propagating in the gauge--invariant two--point function satisfy the Gauss law in presence of the charge density $j_0(\vec{x}) = \sum_f q_f \psi_f^\dag \psi_f(\vec{x})$ of the dynamical degrees of freedom, and the external time--dependent charge density $J_0(t,\vec{x})$, i.e.
\begin{gather}
   \{ \partial_k E_k(\vec{x}) - j_0(\vec{x}) - J_0(t,\vec{x}) \} | \Psi(t) \rangle = 0 \ .
\end{gather}
The evolution of states is governed by a time--dependent non--hermitean Hamiltonian
\begin{gather}
   \mathcal{H}(t) = \mathcal{H}_0 + \imath \int d^3 x \ A_k(\vec{x}) J_k(t,\vec{x}) \ ,
\end{gather}
where $\mathcal{H}_0$ is the standard gauge--invariant Hamiltonian without external current.

Notice that for $z_0 \gg t \gg y_0$, the four--current vanishes exponentially, i.e.
\begin{flalign}
   & J_0(t,\vec{x})
   =
   \frac{q_h}{2L^3} \bigg\{
   e^{- \frac{\pi}{L} (t-y_0)} \sum_{j=1}^3 \cos \tfrac{\pi (x_j-y_j)}{L}
   + e^{- \frac{\pi}{L} (z_0-t)} \sum_{j=1}^3 \cos \tfrac{\pi (x_j-z_j)}{L}
   \bigg\}
   + O(e^{- \frac{3\pi}{L} \Delta t})
   \ , \\
   & J_k(t,\vec{x})
   =
   \frac{q_h}{2L^3} \bigg\{
   e^{- \frac{\pi}{L} (t-y_0)} \sin \tfrac{\pi (x_k-y_k)}{L}
   + e^{- \frac{\pi}{L} (z_0-t)} \sin \tfrac{\pi (x_k-z_k)}{L}
   \bigg\}
   + O(e^{- \frac{3\pi}{L} \Delta t})
   \ .
\end{flalign}
On the one hand, this is a way to see that the leading exponential behaviour of the two--point function is determined by the ground state in the charged sector of the gauge--invariant Hamiltonian $\mathcal{H}_0$. Therefore the mass defined by means of the two--point function in covariant gauge is the correct one. On the other hand, the unphysical exponentials in the external current mimic the contribution of excited states in the long--distance behaviour of the two--point function. For this reason the covariant gauge is not a suitable choice for the extraction of excited states from two-point functions.

\section{Explicit expressions for two--point functions}
\label{sec:correlators}

In this appendix we provide explicit expressions for the two--point functions used in this work, in which fermions have been integrated out. Because of C$^\star$ boundary conditions, the fermion Wick contractions are not the usual ones in terms of the original fields $\psi_f$ and $\bar{\psi}_f$. For instance, the $\psi \psi$ Wick contraction does not vanish. For this reason, we find more convenient to work with the quark--antiquark doublet $\eta_f$ defined in eq.~\eqref{eq:eta}.

The neutral meson operators considered in this work can be easily written in terms of the $\eta_f$ field,
\begin{flalign}
&
   \bar{s} \gamma_5 d + \bar{d} \gamma_5 s
   =
   - \eta_s^T \sigma_1 C \gamma_5 \eta_d
   =
   - \eta_d^T \sigma_1 C \gamma_5 \eta_s
   \ , \\
&
   \bar{s} \gamma_k d - \bar{d} \gamma_k s
   =
   - \eta_s^T \sigma_1 C \gamma_k \eta_d
   =
   \eta_d^T \sigma_1 C \gamma_k \eta_s
   \ .
\end{flalign}
Charged meson operators are written in a similar way,
\begin{flalign}
&
   \bar{S}^I \gamma_5 U^I + \bar{U}^I \gamma_5 S^I
   =
   - \eta_s^T \sigma_1 \Phi_1^I C \gamma_5 \eta_u
   =
   \eta_u^T \sigma_1 \Phi_{-1}^I C \gamma_5 \eta_s
   \ , \\
&
   \bar{S}^I \gamma_5 U^I - \bar{U}^I \gamma_5 S^I
   =
   - \eta_s^T \sigma_1 \sigma_3 \Phi_1^I C \gamma_5 \eta_u
   =
   \eta_u^T \sigma_1 \sigma_3 \Phi_{-1}^I C \gamma_5 \eta_s
   \ , \\
 &
   \bar{S}^I \gamma_k U^I - \bar{U}^I \gamma_k S^I
   =
   - \eta_s^T \sigma_1 \Phi_1^I C \gamma_k \eta_u
   =
   \eta_u^T \sigma_1 \Phi_{-1}^I C \gamma_k \eta_s
   \ ,
\qquad\qquad I=\{\text{s},\text{c}\} \ ,  
\end{flalign}
where $\Phi_{q}^I(x)$ are field--dependent dressing matrices that depend on the choice of the gauge invariant interpolating operator. For string interpolating operators
\begin{gather}
   \Phi_q^\text{s}(x)
   =
   \frac{1}{3} \sum_{k=1}^3 \text{diag} \left(
      \prod_{s=0}^{L-1} U_k(x+s a \hat k)^{-3q}
      \ , \ 
      \prod_{s=0}^{L-1} U_k(x+s a \hat k)^{3q}
   \right)
   \ ,
\end{gather}
while for Coulomb interpolating operators
\begin{gather}
   \Phi_q^\text{c}(x) = \frac{1}{3} \sum_{k=1}^3 \text{diag} \left(
      \prod_{s=0}^{L-1} [ U_k(x+s a \hat k) e^{-\imath A^\text{c}_k(x+s a \hat k)} ]^{-3q}
      \ , \ 
      \prod_{s=0}^{L-1} [ U_k(x+s a \hat k) e^{-\imath A^\text{c}_k(x+s a \hat k)} ]^{3q}
   \right)
   \ .
\end{gather}
Fermionic Wick contractions are generated by the following rule
\begin{gather}
   \bcontraction{}{\eta}{{}_f(x)}{\eta}
   \eta_f(x) \eta_{f'}^T(y) = - \delta_{f,f'} D_f^{-1}(x;y) \sigma_1 C^{-1} \ ,
\end{gather}
where $D_f$ is the $O(a)$--improved Wilson--Dirac operator defined in eq.~\eqref{eq:dirac}.
The relevant mesonic two--point functions are readily calculated. For neutral mesons,
\begin{gather}
   \langle P^0(t) P^0(0) \rangle
   =
   - \frac{1}{4L^3} \sum_{\vec{x}} \langle \tr [ \gamma_5 D_d^{-1}(t,\vec{x};0) \gamma_5 D_s^{-1}(0;t,\vec{x}) ] \rangle
   \ ,
   \\
   \langle V^0_k(t) V^0_k(0) \rangle
   =
   \frac{1}{4L^3} \sum_{\vec{x}} \langle \tr [ \gamma_k D_d^{-1}(t,\vec{x};0) \gamma_k D_s^{-1}(0;t,\vec{x}) ] \rangle
   \ ,
\end{gather}
and similarly for charged mesons with $I=\{\text{s},\text{c}\}$,
\begin{gather}   
   \langle P^I(t) P^I(0) \rangle
   =
   - \frac{1}{4L^3} \sum_{\vec{x}} \langle \tr [ \gamma_5 \Phi_1^I(t,\vec{x}) D_d^{-1}(t,\vec{x};0) \gamma_5 \Phi_{-1}^I(0) D_s^{-1}(0;t,\vec{x}) ] \rangle
   \ , 
   \\
   \langle V^I_k(t) V^I_k(0) \rangle
   =
   \frac{1}{4L^3} \sum_{\vec{x}} \langle \tr [ \gamma_k \Phi_1^I(t,\vec{x}) D_u^{-1}(t,\vec{x};0) \gamma_k \Phi_{-1}^I(0) D_s^{-1}(0;t,\vec{x}) ] \rangle \ .
\end{gather}
We rewrite the interpolating operator for a P$+\gamma$ state in the V channel as
\begin{gather}
   W^I_k(t)
   =
   - \frac{1}{2L^3}
   \sum_{\vec{x}}
   \Xi_k(t,\vec{x}) \,
   \eta_s^T \sigma_1 \sigma_3 \Phi_1^I C \gamma_5 \eta_u(t,\vec{x})
   \ , \\
   \Xi_k(t,\vec{x})
   =
   \sum_{\vec{p} \in O_h \bar{\vec{p}}} e^{-i\vec{p}\vec{x}} \epsilon_{k\ell j} p_\ell \tilde{A}^\text{c}_j(t,\vec{p})
   \ .
\end{gather}
The two new correlators used for the generalised--eigenvalue problem in section~\ref{sec:numericsB} are
\begin{gather}   
   \langle W^I_k(t) V^I_k(0) \rangle
   = \\ \quad =
   \frac{1}{4L^3}
   \sum_{\vec{x}}
   \langle
   \Xi_k(t,\vec{x}) \,
   \tr [ \sigma_3 \Phi_1^I(t,\vec{x}) \gamma_5 D_u^{-1}(t,\vec{x};0)
   \Phi_{-1}^I(0) \gamma_k D_s^{-1}(0;t,\vec{x}) ]
   \rangle
   \ ,  \nonumber \\
   \langle W^I_k(t) W^I_k(0) \rangle
   = \\ \quad =
   \frac{1}{4L^3}
   \sum_{\vec{x}}
   \langle
   \Xi_k(t,\vec{x}) \Xi_k(0) \,
   \tr [ \sigma_3 \Phi_1^I(t,\vec{x}) \gamma_5 D_u^{-1}(t,\vec{x};0)
   \sigma_3 \Phi_{-1}^I(0) \gamma_5 D_s^{-1}(0;t,\vec{x}) ]
   \rangle
   \ . \nonumber
\end{gather}

\bibliographystyle{JHEP}
\bibliography{paper}

\end{document}